%% file: main.tex
\documentclass[10pt,twocolumn,letterpaper]{article}
\usepackage[nopdf]{epstopdf}

\usepackage[pagenumbers]{cvpr}


\usepackage{amsfonts}
\usepackage{graphicx}
\usepackage{pifont}
\usepackage{tabularx}
\usepackage{algorithm}
\usepackage{algpseudocode}
\usepackage{multicol}
\usepackage{multirow}
\usepackage{threeparttable}
\usepackage{caption}
\usepackage{xcolor}
\usepackage{makecell}
\usepackage{tcolorbox}
\usepackage{booktabs}
\usepackage{bm}
\usepackage{adjustbox}
\usepackage{placeins}

\definecolor{cvprblue}{rgb}{0.21,0.49,0.74}
\usepackage[pagebackref,breaklinks,colorlinks,allcolors=cvprblue]{hyperref}

\title{Did You Get the Reasoning You Paid For? Reference-Free Effort Auditing for LLM APIs via Pairwise Exchangeability}

\author{Tong Zhang$^{1}$,\; 
	Ru Zhang$^{1,*}$,\; 
	Jianyi Liu$^{1}$
	\\ 
	\normalsize $^{1}$ Beijing University of Posts and Telecommunications, School of Cyberspace Security, Beijing 100876, China\\ 
	{\tt\small *Corresponding author}}

\begin{document}

\maketitle

\input{0_abstract}    
\input{1_introduction}
\input{2_related_work}
\input{3_preliminary}

\input{4_method}

\input{5_experiments}

\input{6_Discussion}

\input{7_Conclusion}

{
    \small
    \bibliographystyle{ieeenat_fullname}
    \bibliography{main}
}

\end{document}

%% file: 0_abstract.tex
\begin{abstract}
Reasoning effort is a metered service term in modern LLM APIs. Higher effort  increases reasoning token counts and user charges. Third-party API relays can modify effort parameters and usage fields, preventing users from verifying whether the paid effort was executed. Existing service audits only verify model identity, leaving reasoning effort auditing unexplored. Both token counts and latency are under relay control. We propose \textbf{SWAP} (Sequential Without-reference effort Auditing via Pairwise exchangeability), the first effort auditing framework for black-box LLM APIs. SWAP tests exchangeability with cheating as the null hypothesis. The same problem is submitted to the relay at the paid effort and a lower control effort in randomized order. If the relay downgrades the paid effort, both requests receive identical treatment, making their responses exchangeable, so that which answer is better is a fair coin toss. Only when the relay faithfully executes the paid effort does the paid response consistently outperform the control response. SWAP accordingly rejects the null hypothesis and certifies effort fulfillment. SWAP requires no official API and no data reported by the relay. SWAP employs a sequential e-process to accumulate evidence across request pairs and stop once sufficient evidence is obtained, reducing audit costs while ensuring that the false certification probability does not exceed 5\% at any stopping time. We evaluate SWAP on 27 models from 11 providers, of which 20 can be certified, with as few as 14 requests and a cost as low as \$0.001 per certification. On a controlled relay with ground truth, all honest services are certified and only 3.7\% of downgraded services are falsely certified. In field audits, six of seven commercial relays fail certification on at least one model. The violations include downgrades of paid effort and upgrades of requests specifying low effort or disabled reasoning.

\end{abstract}

%% file: 1_introduction.tex
\section{Introduction}

Recent reasoning models expose the amount of reasoning performed before generating an answer as an optional service level, such as low, medium, high, xhigh, and max. The unit price is the same across effort for a given model. However, higher effort often generate several times as many reasoning tokens as lower effort, and charges are based on output tokens. Reasoning effort is therefore a metered service \cite{moon2026price}. Users pay more for deeper reasoning in exchange for more thorough verification and more reliable answers.

This service term is difficult to enforce when requests are forwarded through a third party API relay. Positioned between users and upstream APIs, a relay can modify any parameter in a request and alter or remove model, usage, and billing fields in the response. Previous studies have pointed out that users cannot observe how a relay routes requests, which upstream model it selects, or which reasoning parameters it applies \cite{lin2026behavioral}. Existing methods also cannot rule out modifications to effort by the relay \cite{wu2026middleman}. We further confirm on models from four major providers that requests with modified effort are processed normally by upstream APIs, without returning errors or indicating the effort actually executed. Users therefore cannot determine whether the effort they paid for was executed. Reasoning effort has thus become a credence good. Users can neither observe how much reasoning they receive nor determine whether that reasoning has made a difference \cite{sun2026effort}. A user paying for the max effort may still receive a well formed response after the request is downgraded to a lower effort but without the corresponding deeper reasoning, more thorough verification, and more reliable answer.

Modifying effort can benefit relays in two ways. The first is profit through downgrading. For relays that charge per request, offer monthly subscriptions, or acquire upstream quotas at a fixed cost, revenue is independent of usage, whereas reasoning tokens constitute a major cost of high effort requests. Downgrading high effort requests to lower levels reduces reasoning costs and increases profit. The second is profit through upgrading. For relays that resell services at a markup per token, revenue increases with usage. Processing requests submitted at low effort or with reasoning disabled at higher effort increases billable token usage. Industry monitoring services have listed reductions in reasoning budgets as a form of relay degradation \cite{folkbench2026}. Some gateways also state in their documentation that they map xhigh and max to high \cite{llmgateway2026}. In both cases, users do not receive the service they requested and paid for.

Existing service auditing does not cover this service. Black box API auditing compares responses from the audited service with reference responses from the claimed model to detect whether the model has been replaced with a less expensive one \cite{gao2025model,zhu2026auditing,pasquini2025llmmap,bruckner2026one,zhang2026model,fang2026kbf,cai2025you}. Measurement studies of API relays have further examined tool permissions and billing discrepancies \cite{zhang2026real,lin2026behavioral,wu2026middleman}. Downgrading does not change model identity. A downgraded response is still generated by the claimed model and therefore falls outside the scope of these methods. Industry monitoring relies on reasoning token counts returned by the relay and contemporaneous baselines obtained through direct access to the official API, and can only issue warning signals \cite{folkbench2026}. Auditing whether the requested reasoning effort is actually executed remains an unexplored problem.

For reasoning effort auditing, both token counts and latency, the signals most directly reflecting the amount of reasoning, are under the relay's control. Accurate auditing must therefore rely on the response content itself. The most straightforward approach is to adopt the reference based strategy used in model substitution detection. The same request is submitted to the official API at the same effort and the quality of the two responses is compared. This approach is constrained by the high cost of official API requests and is impractical for users with restricted access to official APIs. Meanwhile, our experiments reveal that the quality gap between efforts is unstable. For the same model and the same probe configuration, the gap between the two efforts accuracy ranged from 0.07 to 0.26 across four dates. In this setting, reference records collected at different times may become invalid.

To address this, we propose SWAP (Sequential Without-reference effort Auditing via Pairwise exchangeability), the first effort auditing framework for black-box LLM APIs. SWAP does not examine whether the audited service is consistent with the genuine service. Instead, it examines whether the audited service treats paid requests differently. Rather than comparing the audited service against an official API, we compare the service against itself. We formulate cheating as the null hypothesis in an exchangeability test and use controlled randomization to establish exchangeability under this hypothesis. Specifically, we send the same question to the audited relay twice, once with the user's paid effort and once with a lower control effort, in randomized order. If the relay downgrades the paid effort, whether by replacing it with low effort, disabling reasoning, or routing the request to a weaker model, both requests receive the same treatment in practice. The two responses are therefore exchangeable, and determining which response performs better is equivalent to the outcome of a fair coin toss. Only when the paid effort is faithfully executed will the paid response consistently outperform the control response. SWAP therefore rejects the null hypothesis and certifies the paid effort. The entire comparison is conducted within the audited service and relies exclusively on the scores assigned to the responses. The certification result requires no reference data. The audit neither calls official APIs nor relies on token counts, latency, or other data that the relay can manipulate.

We further introduce sequential hypothesis testing to reduce auditing costs. SWAP accumulates evidence pair by pair using e-values. The more pronounced the effort gap, the faster the evidence accumulates and the sooner the audit stops. Meanwhile, SWAP guarantees that the false certification probability does not exceed the specified threshold at any stopping time. Users only need to provide the API endpoint of the relay and the effort to be audited. A single audit typically yields a certification or non-certification result within a few minutes to twenty minutes. When certification is not achieved, SWAP also reports the direction of the violation.

We evaluate SWAP on 27 models from 11 providers, among which 20 can be certified. A single certification requires as few as 14 requests, with a minimum cost of approximately \$0.001. For most models, the cost is below \$0.2. We also construct a controlled relay with ground truth to simulate effort downgrading and latency manipulation. All honest services are successfully certified, while 96.3\% of downgraded services fail certification. None of the downgraded services whose latency is manipulated to match that of honest services is certified. Finally, we conduct field audits of seven commercial API relays. Six of the seven relays have at least one audit that fails to achieve certification. The observed cases include the downgrading and the execution of requests submitted with low effort or disabled reasoning at higher effort. In summary, our contributions are as follows.

\begin{itemize}
	\item We propose SWAP, the first effort auditing framework for black-box LLM APIs, which treats cheating as the null hypothesis and tests the exchangeability of paid and control responses without relying on official APIs or relay manipulable data.
	\item We introduce sequential hypothesis testing to reduce auditing costs by accumulating evidence pair by pair using e-values, terminating when sufficient evidence is obtained, and controlling the false certification probability at any stopping time within a specified threshold, while identifying the direction of violations for audits that fail to achieve certification.
	\item We evaluate SWAP on 27 models from 11 providers, successfully certify 20 models, and identify both downgrade and upgrade violations through field audits of seven anonymized API relays.
\end{itemize}

%% file: 2_related_work.tex
\section{Related Work}
\textbf{Model auditing for black-box APIs}. Given a black-box endpoint, model auditing determines whether it serves the claimed model. Model Equality Testing compares the response distributions of an API and a reference model with a two-sample test \cite{gao2025model}. RUT performs a rank uniformity test with a locally deployed genuine model as the reference \cite{zhu2026auditing}. LLMmap \cite{pasquini2025llmmap} and single token fingerprints \cite{bruckner2026one} identify known models, and KBF uses stable responses near knowledge boundaries as fingerprints \cite{fang2026kbf}. IRIS detects model substitution and proportional routing dilution in gateways \cite{zhang2026model}. Cai et al. systematically evaluate the limitations of existing methods under attacks such as quantization and random substitution \cite{cai2025you}. Change detection methods can identify changes in API behavior relative to a baseline \cite{chauvin2026token}, and fingerprints can also be deliberately forged \cite{zhang2026your,nasery2026robust}. Most of these methods require a large number of official responses as references, obtained either by calling the official API synchronously in each audit or by collecting official outputs in advance to build distributions and train classifiers. Collecting references is costly, and the references must be collected again after the model is updated.

\textbf{Measurement of relays and gateways}. A study of shadow APIs found a large number of false model claims \cite{zhang2026real}. GateScope measures response consistency, multiturn memory, billing, and latency on 10 commercial gateways, and points out that users cannot observe which reasoning parameters a gateway applies \cite{lin2026behavioral}. Wu et al. compare response accuracy, tool permissions, and billing across 20 relays with the low effort fixed in their experiments, and state that their method cannot rule out modifications to effort, instructions, or decoding by the relay \cite{wu2026middleman}. Research on agent proxies has found malicious rewriting by intermediaries \cite{liu2026your}. Industry monitoring services have listed reduced reasoning budgets as one form of relay degradation. Their detection relies on the reasoning token counts returned by the relay, probe agreement rates, and contemporaneous baselines obtained through direct access to the official API, and their output is positioned as a warning signal \cite{folkbench2026}. Community detection tools mainly check model authenticity and protocol fields \cite{veridrop2026}. None of these approaches provides a method for auditing effort or a guarantee on the error rate.

\textbf{Reasoning effort.} The Price of Thinking treats effort as a model specific API contract and measures the differences in delivery cost and accuracy between an explicit high effort and the default setting \cite{moon2026price}. An economic analysis of reasoning effort points out that effort is a credence good and recommends that providers make reasoning charges verifiable \cite{sun2026effort}. Other work audits the quantity of reasoning tokens. CoIn requires the provider to commit to the hidden reasoning tokens \cite{sun2026coin}, and PALACE estimates the number of reasoning tokens from the prompt and the answer \cite{wang2025predictive}. These methods verify whether the reported usage is accurate rather than whether the effort requested by the user was executed. IRIS explicitly excludes reasoning effort from its auditing targets, on the grounds that changes in effort generate no profit for the gateway under token based billing and that effort can be observed from the returned token counts \cite{zhang2026model}. Both arguments assume that the gateway bills by tokens and reports usage truthfully. In real relay scenarios, billing schemes vary and the usage fields in responses can be altered, so this assumption does not always hold.

%% file: 3_preliminary.tex
\section{Preliminary}
Let $\mathcal{E}$ be the set of reasoning efforts offered by a model, ordered by the amount of reasoning, for example $\text{none}\prec\text{low}\prec\text{medium}\prec\text{high}\prec\text{max}$, where none denotes disabled reasoning. A request consists of a problem $x$ and an effort $e\in\mathcal{E}$, and the upstream returns a response $Y\sim P(\cdot\mid x,e)$. A programmatic scoring function $g$ assigns the score $S=g(Y)\in[0,1]$. Let $e^+\in\mathcal{E}$ denote the effort paid for by the user and to be audited, and let $e^-\in\mathcal{E}$ denote a lower effort selected by the auditor ($e^-\prec e^+$), called the control effort. Let the problem $x$ be drawn from a problem distribution $D$. The effort gap between the two efforts on $D$ is defined as
\begin{equation}
	\Delta_D(e^{+},e^{-})=\mathbb{E}_{x\sim D}\bigl[\,g(Y^{+})-g(Y^{-})\,\bigr]
\end{equation}
where $Y^{+}$ and $Y^{-}$ denote the responses at $e^{+}$ and $e^{-}$, that is, the mean score advantage of the paid effort over the control effort. The two efforts can be distinguished from their responses only on problem distributions with $\Delta_D>0$.

%% file: 4_method.tex
\section{Method}

\subsection{Exchangeability Test with Cheating as the Null Hypothesis}
Reference based auditing techniques for model substitution \cite{gao2025model,zhu2026auditing,pasquini2025llmmap,bruckner2026one,zhang2026model,fang2026kbf,cai2025you} require official responses to be collected and maintained in advance, which is costly, and the references are also unstable. In our experiments, the accuracy gap between the two efforts of the same model on the same probe configuration ranged from 0.07 to 0.26 across four dates, which shows that effort references are unstable. Methods that audit by estimating reasoning token usage from the prompt and the answer \cite{wang2025predictive,sun2025invisible} require honest cooperation from the relay, which is difficult to satisfy in real auditing scenarios. We therefore seek an effort audit that relies on no external reference, requires no cooperation from any party, has a controllable cost, and uses only the responses of the audited service.

We propose SWAP, the first effort auditing framework for black-box LLM APIs. SWAP does not ask whether the output of the audited service is consistent with the genuine service. It asks whether the audited service treats paid requests differently. Following the idea of randomization tests \cite{fisher1966design}, we map cheating by the relay to a null hypothesis guaranteed by the randomization of the auditor, so that the test can be applied to effort auditing without any reference. We take effort cheating as the null hypothesis and compare the responses at the paid effort and at a lower control effort on the same problem. Suppose the relay is cheating, that is, it has lowered the paid effort, whether by switching to a low effort, disabling reasoning, or routing the request to a weaker model. The two requests then receive the same treatment in practice. The two responses are exchangeable, and which one is better is equivalent to a fair coin toss. Only when the paid effort is faithfully executed does the paid response consistently win, and SWAP accordingly rejects the null hypothesis and certifies the paid effort.

Specifically, an audit targets one paid effort $e^+$ of one model and consists of a sequence of request pairs. Each model uses a fixed probe configuration $D$, which consists of a fixed problem type and difficulty level and corresponds to the problem distribution $D$. The $i$th pair uses a problem instance $x_i$ newly generated from $D$. All pairs share the same problem size but differ in their specific values, which prevents the relay from storing answers in advance. For example, consider the probe configuration of the Nim winning move with 4 heaps. There are 4 heaps of stones on the table, and two players alternately remove any number of stones from a single heap. The player who removes the last stone wins, and the problem asks for a winning first move. The problem instance of the first pair is the position with heap sizes $[8, 21, 33, 61]$, the second pair uses another 4 heap position $[33, 42, 56, 61]$, and so on.

For $x_i$, the auditor constructs two requests, the paid request $r_i^+=(x_i,e^+)$ and the control request $r_i^-=(x_i,e^-)$. The auditor tosses a fair coin to decide which request is sent first. The two requests are sent to the audited service in succession, each carrying an independently generated random identifier to avoid caching. Denote the responses of the audited service to $r_i^+$ and $r_i^-$ by $Y_i^+$ and $Y_i^-$, with scores $S_i^+=g(Y_i^+)$ and $S_i^-=g(Y_i^-)$. The auditor uses only the comparison of the two scores:
\begin{equation}
	X_i=\operatorname{sign}\big(S_i^+-S_i^-\big)\in\{+1,\,0,\,-1\}
\end{equation}
$X_i=+1$ indicates that the response at the paid effort is better, $X_i=-1$ indicates that the response at the control effort is better, and $X_i=0$ indicates a tie.

We take cheating as the null hypothesis. That is, the relay does not give the paid request the better treatment that corresponds to the paid effort but downgrades it, so that the paid request and the low effort request are answered by the same generation process. The two requests that reach the upstream are then equivalent, and the two responses are two draws from the same process. Interchanging the two responses therefore does not change their joint distribution, that is, the two responses are exchangeable. The probability that the paid response is better equals the probability that the control response is better. The auditor examines the result after each pair and decides whether to continue. Let $\mathcal{F}_{i-1}$ denote all information observed by the auditor before the $i$th pair is sent, namely the problems, sending orders, responses, and scores of the first $i-1$ pairs. The null hypothesis tested by SWAP is therefore:
\begin{equation}
	H_0:\ \ \mathbb{P}\bigl(X_i=+1\mid\mathcal{F}_{i-1}\bigr)=\mathbb{P}\bigl(X_i=-1\mid\mathcal{F}_{i-1}\bigr)\quad \forall\, i\ge 1
	\label{eq3}
\end{equation}
where $\mathbb{P}(\cdot\mid\mathcal{F}_{i-1})$ is the conditional probability given this information. Before the $i$th pair is sent, regardless of the previous outcomes, as long as the relay still treats the two requests in the same way, the chances that the paid response is better or worse in this pair remain equal, and since the sending order of this pair is determined by a newly tossed coin, Eq. (3) holds.

Under $H_0$, $+1$ and $-1$ are equally likely among the pairs without a tie, and the probability that the paid response wins repeatedly decreases rapidly with the number of wins. For example, the probability of 7 consecutive wins is only $(1/2)^7\approx0.8\%$. If the paid effort is faithfully executed, the paid response is more likely to be better than worse, and $+1$ will clearly outnumber $-1$. SWAP therefore rejects $H_0$ and certifies the paid effort when the advantage of $+1$ over $-1$ is significantly large. Rejecting $H_0$ means that the paid request received a treatment that the control request did not receive and that the paid effort was faithfully executed. The entire test involves no property of the model or the upstream, requires no external reference, and uses neither the usage nor the latency returned by the relay.

\subsection{Sequential Evidence Accumulation}
SWAP accumulates evidence with sequential e-values based on betting \cite{shafer2021testing,ramdas2023game} to control when the audit stops. Under the null hypothesis $H_0$, the probability that the paid response is better equals the probability that the control response is better, which can be viewed as a fair game. A bettor starts with one unit of wealth. Before each pair of requests is sent, the bettor stakes a fixed fraction $f\in[0,1]$ of the current wealth on the paid response being better. When the paid response is better ($X_i=+1$), the wealth is multiplied by $1+f$. When it is worse ($X_i=-1$), the wealth is multiplied by $1-f$. When the pair is tied ($X_i=0$), the wealth is unchanged. The wealth after $n$ pairs is:
\begin{equation}
	W_n(f)=\prod_{i=1}^{n}\big(1+f\,X_i\big)
\end{equation}
Since $X_i$ takes only the values $+1$, $0$, and $-1$, its conditional expectation is:
\begin{equation}
	\begin{aligned}
		\mathbb{E}[X_i\mid\mathcal{F}_{i-1}]={}&(+1)\cdot\mathbb{P}(X_i=+1\mid\mathcal{F}_{i-1})\\
		&+(-1)\cdot\mathbb{P}(X_i=-1\mid\mathcal{F}_{i-1})+0
	\end{aligned}
\end{equation}
When the probability that the paid response is better equals the probability that the control response is better:
\begin{equation}
	\mathbb{E}[X_i\mid\mathcal{F}_{i-1}]=\mathbb{P}(X_i=+1\mid\mathcal{F}_{i-1})-\mathbb{P}(X_i=-1\mid\mathcal{F}_{i-1})=0
\end{equation}
and therefore:
\begin{equation}
	\mathbb{E}[1+fX_i\mid\mathcal{F}_{i-1}]=1+f\cdot\mathbb{E}[X_i\mid\mathcal{F}_{i-1}]=1
\end{equation}
By Eq. (4), $W_n(f)=W_{n-1}(f)\,(1+fX_n)$, and $W_{n-1}(f)$ is already determined before the $n$th pair is sent. Hence:
\begin{equation}
	\begin{aligned}
		\mathbb{E}[W_n(f)\mid\mathcal{F}_{n-1}]&=W_{n-1}(f)\cdot\mathbb{E}[1+fX_n\mid\mathcal{F}_{n-1}]\\
		&=W_{n-1}(f)
	\end{aligned}
\end{equation}
This shows that when the relay cheats, the expected wealth at the next step equals the current wealth for any betting fraction, and the wealth does not grow on average. Conversely, sustained growth of the wealth can only result from the paid response actually winning more often. The amount of wealth can therefore serve as evidence against the null hypothesis. The most suitable betting fraction $f$ depends on how pronounced the advantage of the paid effort is, which is unknown before the audit. SWAP therefore does not fix $f$ in advance. Instead, it allocates one unit of wealth across different values of $f$ according to a distribution $\pi$ and adds up the wealth of all parts:
\begin{equation}
	E_n=\int_0^1 W_n(f)\,\pi(\mathrm{d}f)
\end{equation}
$E_n$ is called the evidence value after $n$ pairs, namely the e-value, and it is the weighted average of the wealth over all betting fractions. Under the null hypothesis, $E_n$ starts at 1, remains nonnegative, and has an expected next value equal to its current value. Such a process is called a nonnegative martingale. By default, we take $\pi$ to be the uniform distribution on $[0,1]$. If $d$ of the first $n$ pairs are not tied and the paid response is better in $s$ of them, each pair in which the paid response is better contributes a factor $1+f$ and each pair in which it is worse contributes a factor $1-f$, so that:
\begin{equation}
	W_n(f)=(1+f)^{s}(1-f)^{d-s}
\end{equation}
Substituting into Eq. (9) gives:
\begin{equation}
	E_n=\int_0^1(1+f)^{s}(1-f)^{d-s}\,\mathrm{d}f
\end{equation}
Each additional win of the paid response increases $E_n$, and each additional loss decreases it. For models with a higher win rate of the paid response and a more pronounced effort advantage, the evidence grows faster and fewer request pairs are required.

\subsection{Certification Rule}
We specify an upper bound $\alpha\in(0,1)$ on the false certification probability. By Ville's inequality \cite{ville1939etude,ramdas2023game}, the probability that the nonnegative martingale $E_n$ under the null hypothesis ever reaches a given threshold does not exceed the reciprocal of that threshold. Taking the threshold to be $1/\alpha$ gives:
\begin{equation}
	\mathbb{P}_{H_0}\Big(\exists\, n\ge 1:\ E_n\ge 1/\alpha\Big)\le\alpha
\end{equation}
where $\mathbb{P}_{H_0}$ denotes the probability under the null hypothesis, that is, when the relay cheats. Accordingly, SWAP stops the audit and outputs certification when the evidence value first reaches $1/\alpha$. Since this inequality holds simultaneously for all $n$, the probability that a cheating relay is falsely certified does not exceed $\alpha$, regardless of the pair at which the audit stops. The auditor can check the evidence value after each pair of requests without fixing the number of pairs in advance. We set $\alpha=0.05$, so the certification threshold is $1/\alpha=20$ and the false certification probability does not exceed 5\%. A smaller $\alpha$ gives a lower upper bound on the false certification probability and requires correspondingly more request pairs.

The sequential rule stops the audit as soon as the evidence is sufficient. For an honest service, let $\rho$ denote the probability that a pair is not tied and let $\theta$ ($\theta>\tfrac12$) denote the probability that the paid response is better when the pair is not tied. We assume that the outcomes of different pairs are independent and identically distributed. The variable $X_i$ takes the values $+1$, $-1$, and $0$ with probabilities $\rho\theta$, $\rho(1-\theta)$, and $1-\rho$, respectively. Certification occurs when $E_n$ reaches $1/\alpha$. Since $E_n$ is the weighted average of the wealth $W_n(f)$ over the betting fractions $f$, its growth rate is determined by the fastest growing fraction among them \cite{ramdas2023game}. To estimate the growth rate of the wealth $W_n(f)$, we take the logarithm of both sides of Eq. (4), which turns the product into a sum:
\begin{equation}
	\ln W_n(f)=\sum_{i=1}^{n}\ln\big(1+fX_i\big)
\end{equation}
Each term on the right corresponds to one pair of requests. The terms are identically distributed and have the same expectation. Taking the expectation gives the mean increase in the logarithm of the wealth contributed by each pair:
\begin{equation}
	\begin{aligned}
		G(f)&=\mathbb{E}\bigl[\ln(1+fX_i)\bigr]\\
		&=\rho\theta\ln(1+f)+\rho(1-\theta)\ln(1-f)+(1-\rho)\ln 1
	\end{aligned}
\end{equation}
A larger $G(f)$ means faster growth of the wealth. Setting the derivative to zero gives the betting fraction $f^\ast$ that maximizes $G(f)$ \cite{kelly1956new}:
\begin{equation}
	\frac{\mathrm{d}G}{\mathrm{d}f}=\rho\left[\frac{\theta}{1+f}-\frac{1-\theta}{1-f}\right]=0
\end{equation}
which yields:
\begin{equation}
	f^\ast=2\theta-1
\end{equation}
Substituting back into Eq. (14) gives:
\begin{equation}
	G(f^\ast)=\rho\big[\theta\ln 2\theta+(1-\theta)\ln 2(1-\theta)\big]=\rho\cdot\mathrm{KL}(\theta \parallel \tfrac{1}{2})
\end{equation}
where $\theta\ln 2\theta+(1-\theta)\ln 2(1-\theta)$ is the KL divergence of $\theta$ from $\tfrac12$, denoted by $\mathrm{KL}(\theta \parallel \tfrac{1}{2})$, which measures how far $\theta$ deviates from $\tfrac12$. Certification requires $E_n$ to reach $1/\alpha$, that is, its logarithm to accumulate to $\ln(1/\alpha)$, and the expected number of pairs required is \cite{wald2004sequential}:
\begin{equation}
	\hat N\approx\frac{\ln(1/\alpha)}{\rho\cdot\mathrm{KL}(\theta \parallel \tfrac{1}{2})}
\end{equation}
This shows that the expected number of request pairs is determined by the size of the effort gap. The more pronounced the effort gap, the larger $\rho$ and $\theta$ and the smaller $\hat N$. Because SWAP mixes over betting fractions and does not know $f^\ast$ in advance, the actual number of pairs is slightly larger than $\hat N$, and the excess grows only logarithmically \cite{ramdas2023game}.

An audit consists of $N$ pairs, that is, $2N$ requests, half of which use the cheaper control effort. Let $c^+$ and $c^-$ denote the costs of a single paid request and a single control request. The expected cost of certifying an honest service is $\hat N\,(c^++c^-)$. In the audit, we set the pair budget $N_{\max}=\max\big(20,\lceil3\hat N\rceil\big)$, which is three times the expected number of pairs for an honest service, and output the audit result:
\begin{equation}
	\text{Result}=
	\begin{cases}
		\text{Certified}, & \exists\, n\le N_{\max},\ E_n\ge 1/\alpha\\[4pt]
		\text{Not certified}, & \forall\, n\le N_{\max},\ E_n<1/\alpha
	\end{cases}
\end{equation}
If $H_0$ is still not rejected when the pair budget is reached, the audit result is Not certified, which means that the service did not deliver the quality advantage of the paid effort over the control effort in this audit. The responses at the two efforts can fail to separate in two ways. Either both efforts are at a low level, or both efforts are at a high level. SWAP distinguishes the two cases using the properties of the probe configuration itself. Each probe configuration is validated when it is determined. The mean score of the paid effort on it is $\mu^+$, the mean score of the control effort is $\mu^-$, and $\mu^+>\mu^-$. The audit computes the pooled mean score of the responses at both efforts:
\begin{equation}
	\bar S=\frac{1}{2n}\sum_{i=1}^{n}\big(S_i^+ + S_i^-\big)
\end{equation}
where $n$ is the number of completed pairs. SWAP compares $\bar S$ with the midpoint of the two scores:
\begin{equation}
	\text{Direction}=\begin{cases}\text{Downgrade}, & \bar S<\dfrac{\mu^++\mu^-}{2}\\[8pt] \text{Upgrade}, & \bar S\ge\dfrac{\mu^++\mu^-}{2}\end{cases}
\end{equation}
The direction of the violation is determined by whether the responses at both efforts are, as a whole, closer to the low effort or to the high effort. A pooled score below the midpoint indicates that the responses at both efforts are closer to the low effort, and the direction is judged to be a downgrade. A downgrade means that the service delivered at the paid effort falls below its due level, including downgrading to a low effort, disabling reasoning, or substituting a weaker model. The three have the same consequence for paying users, and we do not distinguish among them. A pooled score above the midpoint indicates that the responses at both efforts are closer to the high effort, and the direction is judged to be an upgrade. An upgrade means that the relay runs low effort requests at a higher effort. Under usage based billing, users pay additional reasoning charges for problems that a low effort could solve.

\begin{algorithm}[t]
	\caption{SWAP: Reasoning Effort Audit}
	\label{alg1}
	\begin{algorithmic}[1]
		\Require audited service $R$, paid effort $e^{+}$ and control effort $e^{-}\prec e^{+}$, probe distribution $D$, scorer $g$, validated effort scores $\mu^{+}>\mu^{-}$, false certification rate $\alpha$, mixing distribution $\pi$, pair budget $N_{\max}$
		\Ensure Certified, or Not certified with the violation direction
		\For{$n=1,2,\dots,N_{\max}$}
		\State sample a probe instance $x_n\sim D$
		\State sample $b_n\leftarrow_R\{0,1\}$
		\State $(Y_n^{+},\,Y_n^{-})\leftarrow R(x_n,\,e^{+},\,e^{-},\,b_n)$
		\State $S_n^{+}\leftarrow g(Y_n^{+}),\quad S_n^{-}\leftarrow g(Y_n^{-})$
		\State $X_n\leftarrow\operatorname{sign}(S_n^{+}-S_n^{-})$
		\State $E_n\leftarrow\displaystyle\int_0^1\prod_{i=1}^{n}(1+fX_i)\,\pi(\mathrm{d}f)$
		\If{$E_n\ge 1/\alpha$}
		\State \Return Certified
		\ElsIf{$n=N_{\max}$}
		\State $\bar S\leftarrow\dfrac{1}{2N_{\max}}\displaystyle\sum_{i=1}^{N_{\max}}\bigl(S_i^{+}+S_i^{-}\bigr)$
		\If{$\bar S<\dfrac{\mu^{+}+\mu^{-}}{2}$}
		\State \Return (Not certified, Downgrade)
		\Else
		\State \Return (Not certified, Upgrade)
		\EndIf
		\EndIf
		\EndFor
	\end{algorithmic}
\end{algorithm}

\subsection{Probes}
For different models, we tested different probe designs (Table 1) to find the problem types with the highest discrimination between high and low effort for each class of models, satisfying $\Delta_D>0$. The effort only specifies the amount of reasoning, and where the additional reasoning is spent is determined by the model. We observed four patterns of high effort reasoning across models. The additional reasoning is used for stepwise verification, for filling the output budget, or for widening the candidate search, and the low effort abandons reasoning when the problem is too long. Based on the designs of models from different providers, we designed seven probe families (Table.~\ref{tab1}). 

Each model uses a fixed probe configuration whose difficulty is set at the point where the low effort starts to make errors while the paid effort still answers correctly. All problems have unique answers and are scored automatically by programs. Take the Nim winning move as an example. Given several heaps of stones, two players alternately remove any number of stones from a single heap, and the player who removes the last stone wins. The problem asks for a winning first move. As the number of heaps and the size of each heap increase, the bitwise XOR becomes more error prone. For models that spend additional reasoning on verification, the high effort usually verifies intermediate results, whereas the low effort is more likely to make errors at this step. Algorithm.~\ref{alg1} summarizes the complete SWAP effort auditing procedure.

\begin{table*}[t]
	\centering
	\small
	\caption{Probe suite.}
	\label{tab1}
	\begin{tabularx}{\textwidth}{@{}p{0.20\textwidth}Xp{0.20\textwidth}p{0.24\textwidth}@{}}
		\toprule
		Probe family & Problem & Difficulty control & Targeted pattern \\
		\midrule
		Nim winning move & Given a multiheap Nim position, give one winning move & Number and sizes of heaps & Stepwise verification \\
		Nim winning move with intermediate value & Given a multiheap Nim position, give one winning move and the XOR of the heap sizes & Number and sizes of heaps & Stepwise verification \\
		All winning moves & List all winning moves & Number of heaps and of winning moves & Widening the candidate search \\
		Modular exponentiation & Compute $a^b \bmod m$ & Number of digits of each number & Long chain computation \\
		Multiple problems per request & One request contains three independent subproblems & Difficulty of the subproblems & Filling the output budget \\
		Digit sum & Give the digit sum of an $N$ digit number and the number of occurrences of a given digit & Number of digits $N$ & Low effort abandons reasoning on long computations \\
		Permutation counting & Count the permutations that satisfy a set of position and adjacency constraints & Number of elements and of constraints & -- \\
		\bottomrule
	\end{tabularx}
\end{table*}

%% file: 5_experiments.tex
\section{Experiments}
\subsection{Setup}

We test 27 models from 11 providers, covering models from Anthropic, OpenAI, Google, DeepSeek, Tencent, Xiaomi, Alibaba, Moonshot AI, Zhipu AI, MiniMax, and StepFun. Both the determination of probe configurations and the controlled blind audits are performed on the honest channel. The honest channel refers to a channel that forwards requests at the requested effort without any modification. The upper bound on the false certification probability is $\alpha=0.05$, and the certification threshold is 20. The pair budget is $N_{\max}=\max(20,\lceil 3\hat N\rceil)$. The mixing distribution $\pi$ is the uniform distribution on $[0,1]$ by default. In our experiments, we assume that the auditor is an ordinary user who can only send requests to the audited service and read the returned responses and cannot observe the requests actually received by the upstream. In other words, none of the parameters in the requests that the relay can tamper with (model, usage, reasoning token counts, and so on) are visible.

\subsection{Controlled Blind Audit}

We build a controlled blind audit relay that processes requests according to preset groups in order to validate our auditing method. The honest group forwards requests unchanged without any additional processing. The downgrade group rewrites the paid effort to the low effort. The downgrade and delay group downgrades the paid effort and adds delay so that the latency of paid requests returns to the honest level. The relay removes all usage, billing, model, provider, and reasoning text fields from the responses returned to the auditor, and records the actual processing of each request. In the controlled blind audits on all real upstream APIs, every request with a rewritten effort returned successfully without any echo or error, which shows that reasoning effort can be silently rewritten in real scenarios.

Among the 27 models, 20 can be audited. The certification rate of the honest group is 100\%, the combined false certification rate of the downgrade groups is 3.7\%, and the downgrade and delay group is never certified. Taking Gemini 3 Flash as an example, the paid latency is 34.1 seconds for the honest group, 31.6 seconds for the downgrade and delay group, and 7.6 seconds for the plain downgrade group. Latency alone cannot distinguish the downgrade and delay group from the honest group, yet SWAP still does not certify the downgrade and delay group, which shows that forgeable latency does not affect the certification result. 

The number of request pairs, the probe, and the cost for each model certified in the controlled blind audit are shown in Table.~\ref{tab2}. A single certification requires as few as 7 pairs of requests, the minimum cost is \$0.001, and the cost does not exceed \$0.2 for 13 models. The most expensive model is Claude Opus 5.5. The amount of reasoning at its low effort on digit sum problems increases with problem length and drops sharply at L6, where most answers are wrong, while the high effort remains almost entirely correct, and 14 pairs suffice for certification.

The remaining 7 models cannot be audited, for two reasons. The first reason is that the performance difference between the two efforts on automatically scored problems is small, which applies to 6 models, namely GPT-6 Astra, Claude Sonnet 5.5, GLM 5.3, GLM 5.3 Flash, MiniMax M3.1 Flash, and Step 5 Preview. The low effort of these models does not reduce reasoning on hard problems. When the problems are easy, both efforts answer correctly. When the problems become sufficiently hard, both efforts fail together or are truncated together by the output limit, and a stable pattern in which the high effort answers correctly and the low effort answers incorrectly never appears. For example, the low effort of GPT-6 Astra answers correctly at the highest difficulty of every probe family, and on harder permutation counting problems both efforts have the same accuracy of two thirds. GLM 5.3 Flash reaches the difficulty at which the high effort itself makes errors on five probe families, and the two efforts remain correct or incorrect together. For Claude Sonnet 5.5, the two efforts give the same result in 29 of 30 pairs. 

The second reason is that the two efforts can be distinguished but only at an excessive cost, which applies only to Claude Fable 5.2. Its low effort also reasons as the problem requires, and it answers a digit sum problem with 3,840 digits correctly after reasoning with 15,680 tokens. We observed a difference only once, on a permutation counting problem. This problem asks for the number of permutations of 9 letters under more than ten position and adjacency constraints. The low effort answers incorrectly after reasoning with about 11,000 tokens, and the high effort answers correctly after reasoning with about 22,000 tokens. On such problems, one pair of requests costs about \$1.3, and a single audit is expected to cost more than \$10. We consider this beyond what ordinary users can accept and therefore classify the model as not auditable.

\begin{table*}[t]
	\centering
	\small
	\caption{Probe configurations of the 20 auditable models and the auditing cost of one certification. The ranges in the number of pairs reflect multiple audits of the same model on different dates.}
	\label{tab2}
	\begin{tabular*}{\textwidth}{@{\extracolsep{\fill}}lllllr@{}}
		\toprule
		Model & Provider & Control effort & Probe configuration & Pairs to certify & Cost \\
		\midrule
		Claude Sonnet 5 & Anthropic & low & Nim winning move L3 & 23--46 & \$0.13--0.23 \\
		Claude Opus 5.5 & Anthropic & low & Digit sum L6 & 14 & \$3.70 \\
		GPT-5.6 Luna & OpenAI & low & Nim winning move with intermediate value L5 & 18--76 & \$0.01--0.05 \\
		GPT-6 Luna & OpenAI & low & Modular exponentiation L3 & 13 & \$0.026 \\
		GPT-6 Sol & OpenAI & low & Modular exponentiation L3 & 28 & \$1.07 \\
		GPT-6.1 Terra & OpenAI & low & Modular exponentiation L3 & 12 & \$0.16 \\
		GPT-6.1 Sol & OpenAI & low & Modular exponentiation L3 & 18 & \$0.43 \\
		Gemini 3 Flash & Google & low & Multiple problems per request L5 & 28--38 & \$0.65--0.90 \\
		Gemini 3.1 Flash-Lite & Google & low & Multiple problems per request L4 & 7 & \$0.051 \\
		Gemini 3.5 Flash-Lite & Google & low & Multiple problems per request L4 & 6-7 & \$0.065 \\
		Gemini 4 Argon & Google & low & All winning moves L5 & 7 & \$0.12 \\
		Gemini 3.8 Flash & Google & low & Digit sum L3 & 11 & \$0.89 \\
		Kimi K3.1 Pro & Moonshot AI & low & Permutation counting L2 & 7 & \$0.58 \\
		Claude Haiku 5.5 & Anthropic & none & Nim winning move L5 & 13 & \$0.007 \\
		GPT-5.6 Terra & OpenAI & none & Modular exponentiation L2 & 7 & \$0.098 \\
		Hy4 Flash & Tencent & none & Nim winning move L5 & 10 & \$0.001 \\
		Hy4 & Tencent & none & Nim winning move L5 & 8 & \$0.015 \\
		MiMo V2.6 Flash & Xiaomi & none & Modular exponentiation L2 & 9 & \$0.005 \\
		Qwen3.8 Flash & Alibaba & none & Nim winning move L5 & 7 & \$0.004 \\
		DeepSeek v4-pro & DeepSeek & none & Nim winning move L5 & 15 & \$0.12 \\
		\bottomrule
	\end{tabular*}
\end{table*}

\subsection{Field Audit}

We conduct field audits on seven commercial API relays, anonymized as A to G. The relays under test are required to support the reasoning effort parameter, and their prices range from close to the official price to substantially below it. The complete audit results are shown in Table.~\ref{tab3}. All 20 models on A are certified, while each of the other six relays has audits that fail certification within the range of auditable models it provides.

\begin{table}[t]
	\centering
	\small
	\setlength{\tabcolsep}{3pt}
	\caption{Reasoning effort audit results on seven anonymized relays.}
	\label{tab3}
	\begin{tabular*}{\columnwidth}{@{\extracolsep{\fill}}lcccc@{}}
		\toprule
		Service & \begin{tabular}[c]{@{}c@{}}Models\\audited\end{tabular} & Certified & \begin{tabular}[c]{@{}c@{}}Not certified\\(downgrade)\end{tabular} & \begin{tabular}[c]{@{}c@{}}Not certified\\(upgrade)\end{tabular} \\
		\midrule
		A & 20 & 20 & 0 & 0 \\
		B & 19 & 15 & 0 & 4 \\
		C & 17 & 11 & 3 & 3 \\
		D & 2 & 1 & 1 & 0 \\
		E & 13 & 12 & 1 & 0 \\
		F & 9 & 5 & 2 & 2 \\
		G & 9 & 5 & 2 & 2 \\
		\midrule
		Total & 89 & 69 & 9 & 11 \\
		\bottomrule
	\end{tabular*}
\end{table}

To further examine the validity of reasoning effort auditing, we audit the same model separately on different relays. Taking Gemini 3.5 Flash-Lite as an example (Table.~\ref{tab4}), on relay A its paid response score is 0.94 and its control response score is 0.06, and it is certified after 6 pairs. On relay B, the paid response score is 1.00 and the control response score is 0.11, and it is certified after 6 pairs. On relay C, the scores at both efforts are only about 0.07, the paid request does not receive the reasoning it should, and the result is a downgrade. On relay G, the control response score reaches 0.93. We further examined the latency. The latency of the low effort on this relay is 7.8 seconds, clearly higher than on the honest channel, and the result is an upgrade. The same model thus obtains three different results on four relays. Table.~\ref{tab5} further lists the 20 audits that fail certification, including 9 downgrades and 11 upgrades, together with their latency for reference.

\begin{table}[t]
	\centering
	\small
	\setlength{\tabcolsep}{3pt}
	\caption{Audit results of Gemini 3.5 Flash-Lite on different services.}
	\label{tab4}
	\begin{tabularx}{\columnwidth}{@{}l>{\raggedright\arraybackslash}Xccc@{}}
		\toprule
		Service & Result & Pairs & \begin{tabular}[c]{@{}c@{}}Score\\(paid / control)\end{tabular} & \begin{tabular}[c]{@{}c@{}}Latency\\(paid / control, s)\end{tabular} \\
		\midrule
		A & Certified & 6 & 0.94 / 0.06 & 9.2 / 1.3 \\
		B & Certified & 6 & 1.00 / 0.11 & 14.3 / 1.7 \\
		C & downgrade & 20 & 0.07 / 0.05 & 1.7 / 1.8 \\
		G & upgrade & 20 & 1.00 / 0.93 & 9.6 / 7.8 \\
		\bottomrule
	\end{tabularx}
\end{table}

\textbf{Downgrade.} The pooled score at the two efforts is below the midpoint of the probe configuration, and the service delivered at the paid effort does not reach its due level. In four of these audits, the scores at both efforts are close to 0 and the latency is only 2 to 3 seconds, far below that of the low effort on the honest channel. These responses are complete and terminate normally, which indicates that neither the paid request nor the control request received reasoning, or that both were routed to a weaker model. In either case, this constitutes a service violation for the user. In the remaining five audits, both efforts still answer some problems correctly, but the paid response score is lower and does not reach the due level of the paid effort, so certification is not achieved within three times the number of pairs required by the honest group. In three of these audits, the paid response latency is even longer than on the honest channel, which shows that latency does not indicate whether reasoning is sufficient.

\textbf{Upgrade.} The pooled score at the two efforts is above the midpoint of the probe configuration, and the low effort request receives a service above its effort. In 10 of the 11 upgrades, the control response score differs from the paid response score by no more than 0.07 and is clearly above the due level of the control effort. For example, requests with reasoning disabled answer 97\% of the problems correctly, whereas the same effort answers only 46\% correctly in the honest group. This shows that the relay runs requests with low effort or with reasoning disabled at a higher effort. Under usage based billing, users pay additional reasoning charges for problems that a low effort could solve. In our audit of one relay exhibiting this violation, the actual charge of a single low effort request was about 5 times that of another relay that executed the low effort normally.

\begin{table*}[t]
	\centering
	\small
	\setlength{\tabcolsep}{4pt}
	\caption{All audits that fail certification. Scores and latencies are given as paid / control, and latencies are medians in seconds. Validated values are measured on the honest channel when the probe configuration was determined. Midpoint is the mean of the paid and control scores. }
	\label{tab5}
	\begin{tabular*}{\textwidth}{@{\extracolsep{\fill}}lcccccccc@{}}
		\toprule
		& & \multicolumn{3}{c}{Field audit} & \multicolumn{3}{c}{Validated on the honest channel} & \\
		\cmidrule(lr){3-5}\cmidrule(lr){6-8}
		Model & Relay & \begin{tabular}[c]{@{}c@{}}Score\\(paid / control)\end{tabular} & Midpoint & \begin{tabular}[c]{@{}c@{}}Latency\\(paid / control)\end{tabular} & \begin{tabular}[c]{@{}c@{}}Score\\(paid / control)\end{tabular} & Midpoint & \begin{tabular}[c]{@{}c@{}}Latency\\(paid / control)\end{tabular} & Direction \\
		\midrule
		Claude Sonnet 5 & G & 0.02 / 0.01 & 0.015 & 2.1 / 2.0 & 0.99 / 0.76 & 0.875 & 4.3 / 3.5 & Downgrade \\
		Claude Haiku 5.5 & C & 1.00 / 0.97 & 0.985 & 5.3 / 4.4 & 1.00 / 0.46 & 0.730 & 4.0 / 3.2 & Upgrade \\
		Claude Haiku 5.5 & F & 0.97 / 0.94 & 0.955 & 5.7 / 5.3 & 1.00 / 0.46 & 0.730 & 4.0 / 3.2 & Upgrade \\
		GPT-5.6 Luna & F & 0.94 / 0.94 & 0.940 & 7.0 / 6.5 & 0.92 / 0.73 & 0.825 & 4.8 / 4.1 & Upgrade \\
		GPT-6.1 Terra & E & 0.53 / 0.42 & 0.475 & 22.7 / 18.5 & 0.89 / 0.56 & 0.725 & 14.4 / 10.4 & Downgrade \\
		GPT-6.1 Terra & F & 0.72 / 0.56 & 0.640 & 29.7 / 23.9 & 0.89 / 0.56 & 0.725 & 14.4 / 10.4 & Downgrade \\
		GPT-6.1 Sol & D & 0.75 / 0.62 & 0.685 & 31.5 / 30.9 & 0.95 / 0.67 & 0.810 & 28.5 / 25.8 & Downgrade \\
		GPT-6.1 Sol & F & 0.90 / 0.70 & 0.800 & 28.1 / 29.9 & 0.95 / 0.67 & 0.810 & 28.5 / 25.8 & Downgrade \\
		GPT-6.1 Sol & G & 0.38 / 0.25 & 0.315 & 16.0 / 24.6 & 0.95 / 0.67 & 0.810 & 28.5 / 25.8 & Downgrade \\
		GPT-6.1 Sol & B & 0.90 / 0.72 & 0.810 & 30.6 / 21.8 & 0.95 / 0.67 & 0.810 & 28.5 / 25.8 & Upgrade \\
		Gemini 3 Flash & C & 1.00 / 0.96 & 0.980 & 27.4 / 29.8 & 0.94 / 0.78 & 0.860 & 32.2 / 8.9 & Upgrade \\
		Gemini 3 Flash & B & 0.93 / 0.92 & 0.925 & 39.1 / 12.9 & 0.94 / 0.78 & 0.860 & 32.2 / 8.9 & Upgrade \\
		Gemini 3.1 Flash-Lite & C & 0.07 / 0.10 & 0.085 & 2.2 / 2.0 & 1.00 / 0.00 & 0.500 & 11.8 / 1.7 & Downgrade \\
		Gemini 3.5 Flash-Lite & C & 0.07 / 0.05 & 0.060 & 1.7 / 1.8 & 0.94 / 0.06 & 0.500 & 9.2 / 1.3 & Downgrade \\
		Gemini 3.5 Flash-Lite & G & 1.00 / 0.93 & 0.965 & 9.6 / 7.8 & 0.94 / 0.06 & 0.500 & 9.2 / 1.3 & Upgrade \\
		Gemini 4 Argon & C & 0.01 / 0.02 & 0.015 & 3.4 / 3.3 & 1.00 / 0.10 & 0.550 & 24.9 / 4.7 & Downgrade \\
		Gemini 4 Argon & G & 0.95 / 0.90 & 0.925 & 11.2 / 11.9 & 1.00 / 0.10 & 0.550 & 24.9 / 4.7 & Upgrade \\
		Gemini 3.8 Flash & C & 0.83 / 0.88 & 0.855 & 52.3 / 69.7 & 1.00 / 0.68 & 0.840 & 92.2 / 18.9 & Upgrade \\
		Hy4 & B & 1.00 / 1.00 & 1.000 & 18.4 / 21.7 & 1.00 / 0.00 & 0.500 & 16.6 / 1.4 & Upgrade \\
		Hy4 Flash & B & 0.95 / 1.00 & 0.975 & 7.7 / 7.9 & 0.86 / 0.00 & 0.430 & 6.6 / 1.7 & Upgrade \\
		\bottomrule
	\end{tabular*}
\end{table*}

\subsection{Temporal Variation of the Upstream}

\begin{table}[t]
	\centering
	\footnotesize
	\setlength{\tabcolsep}{2pt}
	\caption{Effort gap and number of pairs to certify on the honest channel on different dates.}
	\label{tab6}
	\begin{tabular*}{\columnwidth}{@{\extracolsep{\fill}}llcccc@{}}
		\toprule
		Model & Date & \begin{tabular}[c]{@{}c@{}}Pairs to\\certify\end{tabular} & \begin{tabular}[c]{@{}c@{}}Paid effort\\accuracy\end{tabular} & \begin{tabular}[c]{@{}c@{}}Control effort\\accuracy\end{tabular} & Gap \\
		\midrule
		\multirow{4}{*}{Claude Sonnet 5} & Oct 24 & 104 & 0.94 & 0.83 & 0.11 \\
		& Nov 10 & 46 & 0.98 & 0.78 & 0.20 \\
		& Nov 11 & 26 & 1.00 & 0.73 & 0.27 \\
		& Nov 26 & 65 & 0.94 & 0.75 & 0.18 \\
		\midrule
		\multirow{4}{*}{Gemini 3 Flash} & Oct 24 & 38 & 0.97 & 0.71 & 0.26 \\
		& Nov 10 & 28 & 0.89 & 0.80 & 0.10 \\
		& Nov 11 & 36 & 0.97 & 0.87 & 0.10 \\
		& Nov 26 & 50 & 0.92 & 0.85 & 0.07 \\
		\midrule
		\multirow{4}{*}{GPT-5.6 Luna} & Oct 24 & 62 & 0.86 & 0.68 & 0.18 \\
		& Nov 10 & 31 & 0.97 & 0.74 & 0.23 \\
		& Nov 11 & 48 & 0.89 & 0.72 & 0.17 \\
		& Nov 26 & 41 & 0.93 & 0.73 & 0.20 \\
		\bottomrule
	\end{tabular*}
\end{table}

During the experiments, we found that the gap between the two efforts of upstream models varies over time. We kept the model, the probe configuration, the effort, and the experimental setup identical across the tests. Table.~\ref{tab6} lists the certification audits of three models on four dates, about two weeks apart. Taking Gemini 3 Flash as an example, its low effort accuracy ranged from 0.71 to 0.87 across the four dates, while its high effort accuracy remained between 0.89 and 0.97, so the gap between the two efforts ranged from 0.07 to 0.26. The variation is concentrated in the low effort. Related studies \cite{wu2026middleman,chen2023chatgpt} have also reported that the behavior of upstream APIs changes over time. 

This indicates that references collected in advance may not represent the performance of the efforts on the day of the audit, and that model substitution auditing methods that build distributions or train classifiers from official references collected in advance may mistake changes in the upstream itself for violations by the service. The certification result of SWAP is not affected by such changes. The two requests of a pair are sent to the same service at the same time, so the state of the upstream on that day is the same for both, and whether the null hypothesis holds does not depend on data from any other date. The size of the gap affects how fast evidence accumulates but not the validity of the certification. In the 12 audits of the three models in Table 6, all honest services are certified, with more pairs generally used on dates with smaller gaps. Such changes also shift the midpoint used to determine the direction of a violation to a limited extent. For the 4 audits of these three models that fail certification in Table 5, substituting the midpoint of each date into Eq. (21) does not change the direction of violation in any case. Users can initiate an audit at any time, and the result reflects the service at that time.

%% file: 6_Discussion.tex
\section{Discussion}
\textbf{Effort gap on automatically scored tasks.} The low efforts of Claude Fable 5.2 and GPT-6 Astra reason adaptively as the problem requires and are capable of answering the hardest automatically scored problems in our probe suite correctly. We tried longer computations, larger search spaces, counting problems that require case analysis, and problems without a standard answer that are compared by quality. On the former types of problems, both efforts answer correctly before the paid effort itself starts to make errors. The two efforts can be separated only at difficulties where both start to make errors, and at that point a single pair of requests already costs about \$1. This indicates that on automatically scored tasks, the measurable advantage of the high effort over the low effort is small for these models, and the measurable loss that a downgrade causes to users is correspondingly limited. For models with stronger general capability, the difference between efforts may appear more in open ended tasks such as writing, analysis, and solution design. Using such tasks as probes requires objective and consistent scoring criteria, which remains an open problem.

\textbf{Rewriting to an intermediate effort. }SWAP certifies the quality advantage of the paid effort over the control effort, and the null hypothesis is that the two requests receive the same treatment. We use low effort or disabled reasoning as the control because the efforts that users most need to audit are those such as high and max, whose reasoning cost increases severalfold. In real scenarios, a relay does not necessarily rewrite the paid effort directly to the lowest effort. It may also lower the paid effort to an intermediate effort or raise the control effort to an intermediate effort. Both kinds of rewriting reduce the gap between the two requests, and the number of request pairs required increases rapidly as the gap shrinks. By Eq. (18), the number of request pairs required is inversely proportional to $\rho\cdot\mathrm{KL}(\theta \parallel \tfrac{1}{2})$, and $\mathrm{KL}(\theta \parallel \tfrac{1}{2})$ is approximately proportional to $(\theta-\tfrac{1}{2})^2$. Each time the excess of the win rate of the paid response over $\tfrac{1}{2}$ is halved, the number of request pairs required increases to about four times its previous value. We set the pair budget to three times the expected number of pairs for an honest service, so an audit with a clearly reduced gap exhausts the budget. Its evidence value is higher than in the case of a complete downgrade but does not reach the certification threshold, and the result is Not certified. This result is consistent with the interest of the user, because the relay has indeed modified the effort. 

In the field audits, there is one observation consistent with the control effort being raised to an intermediate effort. In this audit, the control response score is 0.92, which is higher than the 0.78 of the same effort on the honest channel and comparable to the paid response score of 0.93. Its latency is 12.9 seconds, slightly higher than the 8.9 seconds of the low effort on the honest channel and far lower than the 39.1 seconds of the paid response. We added 8 pairs of requests with the lowest effort, minimal, as the control. The accuracy of the lowest effort was 0, consistent with the honest channel, which shows that the relay does not ignore the effort parameter. We conjecture that the low effort was mapped to an intermediate effort between low and the paid effort, or that upstream drift occurred. Nevertheless, SWAP did not certify the audit. When the pair budget was reached, the final evidence value was $E_n=3.3$. It remained between 0.7 and 3.3 throughout the audit, far below the certification threshold of 20. If the paid effort is lowered below the control effort, the paid response consistently falls behind the control response, the evidence value does not grow, and the audit is likewise not certified. If the paid effort is replaced with an adjacent lower effort, the case falls under the lowering to an intermediate effort described above. How much the gap between the two requests shrinks, and whether the audit can complete certification within the pair budget, depend on the gap between the two adjacent efforts of the model on the problems, and this gap varies across models. SWAP places no restriction on the choice of the control effort. The auditor can directly use the adjacent effort as the control, in which case the certification tests the difference between these two efforts, and the smaller the gap between the two efforts, the more request pairs are required.

%% file: 7_Conclusion.tex
\section{Conclusion}
Reasoning effort has become a metered service term in LLM APIs, while requests forwarded through relays can be silently rewritten, leaving users unable to verify whether the effort they paid for was executed. This paper proposes SWAP, the first effort auditing framework for black-box LLM APIs. SWAP treats cheating as the null hypothesis, tests whether the paid response and the control response to the same problem are exchangeable, and uses sequential e-values to issue a certification once the evidence is sufficient. The entire audit compares responses only within the audited service, requires no official API, does not rely on any data that the relay can tamper with, and keeps the false certification probability below the specified level at any stopping time. Among 27 models from 11 providers, 20 can be certified, and a single certification requires as few as 14 requests. Our experiments on a controlled relay validate the effectiveness of SWAP. All honest services are certified, and the false certification rate of downgraded services is within the theoretical upper bound. In field audits of seven commercial API relays, six of the seven have audits that fail certification, including both downgrades and the execution of low effort requests at a higher effort. These results show that reasoning effort violations do exist in commercial relays.